\RequirePackage{lineno}
\documentclass[prl,twocolumn,showpacs,superscriptaddress,amsmath,amssymb,nofootinbib]{revtex4-2}
\usepackage{graphicx}
\usepackage{epsfig}
\usepackage{overpic}
\usepackage{dcolumn}
\usepackage{ulem}
\usepackage{bm}
\usepackage{lineno}
\usepackage{xspace}
\usepackage{multirow}
\usepackage{epstopdf}
\usepackage{xcolor}
\usepackage{soul}
\usepackage[colorlinks,allcolors=black]{hyperref}
\usepackage{verbatim}
\usepackage{enumitem}
\usepackage{todonotes}

\usepackage{subfigure}
\include{def-com}

\begin{document}
\title{\boldmath Discover the Gell-Mann–Okubo formula with machine learning}

\author {Zhenyu Zhang}
\address{Guangdong Provincial Key Laboratory of Nuclear Science, Institute of Quantum Matter, South China Normal University, Guangzhou 510006, China}
\affiliation{Guangdong-Hong Kong Joint Laboratory of Quantum Matter, Southern Nuclear Science Computing Center, South China Normal University, Guangzhou 510006, China}

\author{Rui Ma}
\address{Guangdong Provincial Key Laboratory of Nuclear Science, Institute of Quantum Matter, South China Normal University, Guangzhou 510006, China}
\affiliation{Guangdong-Hong Kong Joint Laboratory of Quantum Matter, Southern Nuclear Science Computing Center, South China Normal University, Guangzhou 510006, China}

\author{Jifeng Hu}\email{hujf@m.scnu.edu.cn}
\address{Guangdong Provincial Key Laboratory of Nuclear Science, Institute of Quantum Matter, South China Normal University, Guangzhou 510006, China}
\affiliation{Guangdong-Hong Kong Joint Laboratory of Quantum Matter, Southern Nuclear Science Computing Center, South China Normal University, Guangzhou 510006, China}

\author{Qian Wang}\email{qianwang@m.scnu.edu.cn}
\address{Guangdong Provincial Key Laboratory of Nuclear Science, Institute of Quantum Matter, South China Normal University, Guangzhou 510006, China}
\affiliation{Guangdong-Hong Kong Joint Laboratory of Quantum Matter, Southern Nuclear Science Computing Center, South China Normal University, Guangzhou 510006, China}
\date{\today}

\begin{abstract}

Machine learning is a novel and powerful technology
and has been widely used in various 
science topics. We demonstrate a machine-learning-based approach 
built by a set of general metrics and rules inspired by physics.
Taking advantages of physical constraints, such as dimension identity, symmetry and generalization, 
we succeed to re-discover the Gell-Mann–Okubo formula using a technique of symbolic regression.
This approach can effectively find explicit solutions among user-defined observable, and easily extend to
study on exotic hadron spectrum.

\end{abstract}

\maketitle

{\bf \color{gray}Introduction:}
Spectroscopy is a complex art, but interesting and helpful, 
to shed light on the underlying physics. It has exhibited its power
in cosmology, molecular physics, atomic physics, particle physics and so on.
For instance, in cosmology, it can tell how an object like a black hole,
neutron star, or active galaxy is producing light, how it moves, and even 
the elements of the interested object. In atomic physics, the Rydberg formula
of the spectrum of hydrogen leads to the quantum mechanics interpretation. 
In particle physics, Gell-Mann and Okubo developed eightfold way 
classification schemes for bloomy hadrons uncovered by new experimental 
techniques in 1960s~\cite{Gell-Mann:1961omu,Okubo:1961jc,Gell-Mann:1962yej}. This discovery also 
indirectly lead the establishment of quark model,
which stated that baryon and meson were made of three quarks
and quark-antiquark, respectively. This picture
was challenged until the first observation of exotic 
candidate $X(3872)$ in 2003. Subsequently, numerous exotic
candidates emerged in various high energy accelerators~\cite{Chen:2016qju,Liu:2019zoy,Chen:2016spr,Dong:2017gaw,Lebed:2016hpi,Guo:2017jvc,Albuquerque:2018jkn,Yamaguchi:2019vea,Guo:2019twa,Brambilla:2019esw}. Particle physicists keep trying to discover
a mass formula for describing the observed 
exotic candidates and predicting the new ones. 

Obviously, these work require physicists to develop an excellent
insight and a rich imagination after a long professional training.
Very recently machine learning has been widely used in various topics~\cite{paper:symm}~\cite{paper:gravitation}, 
which has exhibited great success and power in solving specific application problems.
However, it is still at infancy stage in discovering general scientific laws~\cite{paper:scie1} from experimental data. 
Recent activities have been made with different approaches, for example, neural network in finding conserved quantities in gravitation~\cite{paper:cons_law} and the 
equation of a primordial state of matter in high-energy
heavy-ion collisions~\cite{paper:eos} or
symbolic regressions in extracting physical parameters from complex data sets~\cite{paper:symb2} and unknown functions from the Feynman Lectures on Physics~\cite{paper:symb3}.

In this work, we demonstrate a physics-inspired machine-learning framework 
consisting of a set of general rules and metrics to discover
explicit connections from many observable,
even to discover hidden physical laws and to explain more complex systems without too much prior knowledge of particle physics.
More specifically, we establish a machine-learning framework
to find the Gell-Mann-Okubo formula from baryon decuplet and octet. 
Unlike purely mathematical approaches, this approach
takes advantages of physical constraints, such as dimension identity, symmetry and generalization and so on,
to effectively find proper solutions. 

\vspace{0.2cm}
{\bf \color{gray} Framework:}
The primitive framework is described as below:
\begin{enumerate}[labelsep=0.2em, leftmargin=10pt, itemindent=0em]
\item Input data: the observable $\bm{x} \in R^m$ and the target $\bm{y} \in R$ can be a scalar or a vector, for example the motion speed, a position, economic indices of the society and so on. Data in the observation space can be collected from experiments or generated with simulations. Unlike massive data used in deep learning, a few data should be enough. In our case, the decuplet and octet baryon masses are the input data. 

\item Model evolution~\cite{paper:evo1}: 
technically, we use a chain tree to construct an evolutionary model $f$ as illustrated in Fig.~\ref{fig:chaintree1},
in which green (blue) nodes represents operators (observable).
$f$ represents an analytical formula consisting of a set of operators and observable predefined as listed in Tab.~\ref{tab:opobs}, and \textbf{p} free parameters. 
Any operator must carry two observable, one of them can be unfolded into a sub tree until reaching the tail. 
A model is expressed by sequentially folding nodes of a chain tree from right to left.
For instance, Fig.~\ref{fig:chaintree1} (top) represents $f(Y,I)=p_1 + p_2Y$.
Here, $p_1$ and $p_2$ are free parameters, $Y$ and $I$ are observable.
Clearly one operator and two observable can only construct simplest models.
Complicated models stem from evolution of simple models by forking the tail node 
or mutating any node for many times following an idea of the evolutionary algorithm~\cite{paper:evo1}.
Figure~\ref{fig:chaintree1} (bottom) represents another model $f(Y,I)=p_1 + p_2(Y+p_4I)$ after two times evolution.

\begin{figure}[h]
    \centering
    \includegraphics[width=7.2cm,height=2.4cm]{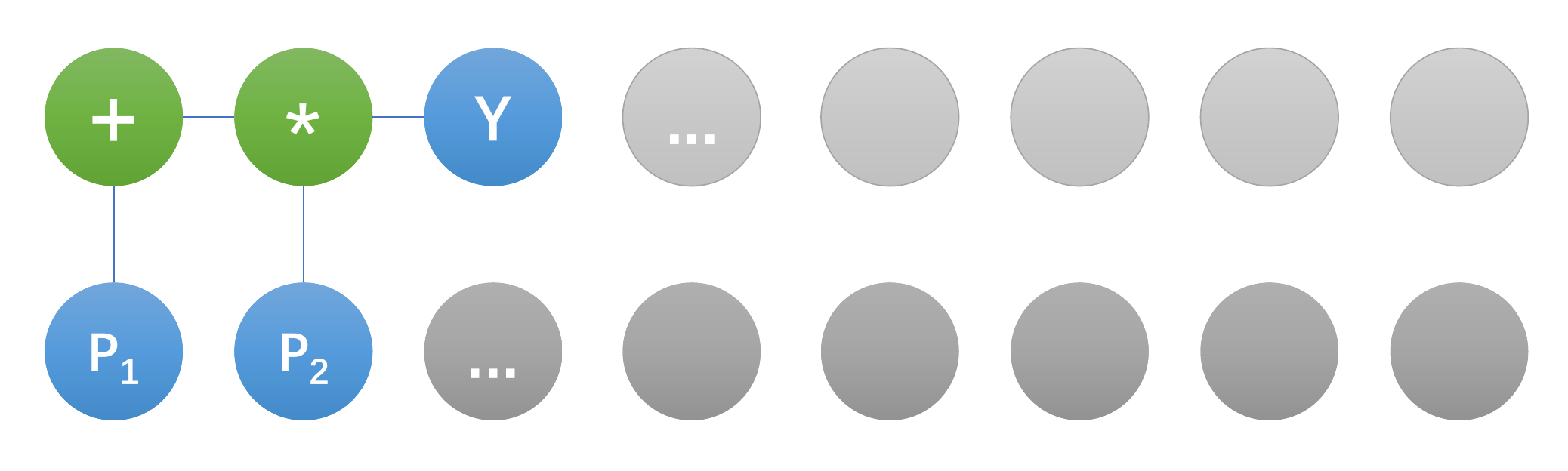}
    \includegraphics[width=7.2cm,height=2.4cm]{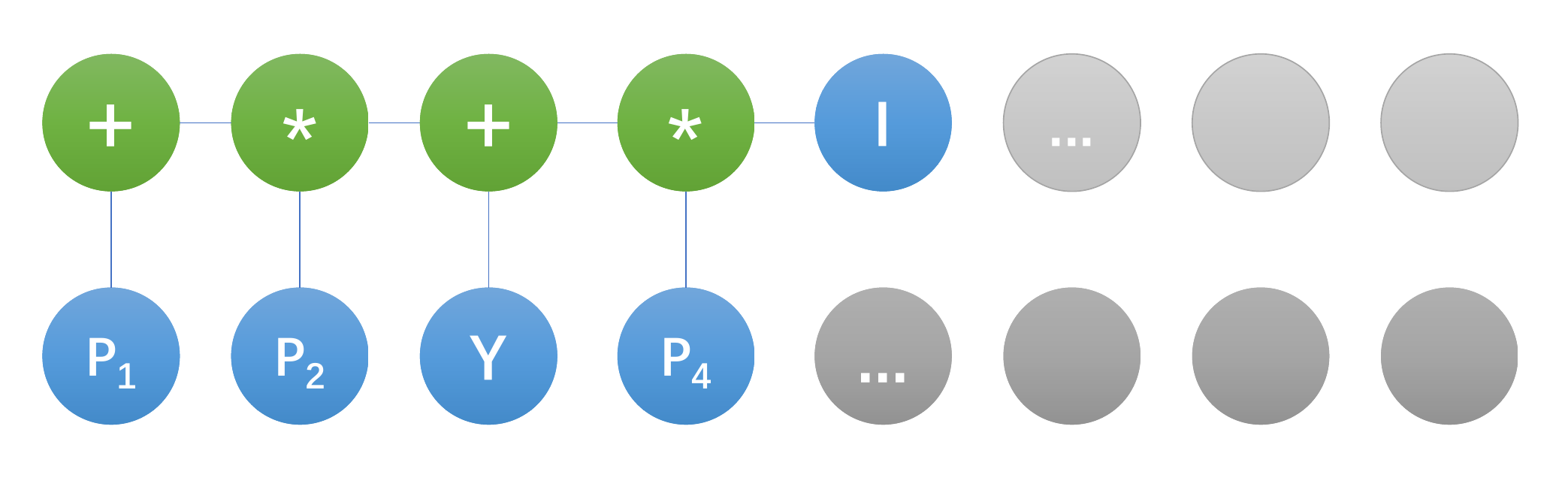}
    \caption{A formula is unfolded from left to right into a tree consisting of a set of operators and a set of observable.
    This formula is calculated by folding the tree from right to left.}
    \label{fig:chaintree1}
\end{figure}

\begin{table}[h]
    \centering
        \caption{predefined operators and observable which are arbitrary in principle.}
    \begin{tabular}{c|c |c |c}
    \hline\hline
         operators & descriptions & observable & descriptions \\
         \hline
         + &  addition  & M & hadron mass \\
         - & subtraction & Y & hypercharge \\
         * & multiplication & I & isospin \\
         / & division & \textbf{p} & free parameters\\
         \hline
    \end{tabular}
    \label{tab:opobs}
\end{table}

\item Physics Filter: 
the challenging task is to construct all models, because available models in mathematical space are infinite. 
Initial enumeration generates all possible models by mutating the operator (observable) with the rest ones given the depth of a chain tree. However, physics filters can strongly guide how to remove meaningless models.
For example, dimension identity requires any two observable belonging to the ``addition" operator of having identical dimension.
Or on the other hand, observable with different dimension only allow for multiplication or division. 
Obviously meaningless formulas $Y/Y$ or $Y-Y$ can be easily filtered out.
These rules are referred to as physical filters.

\item Metrics: they measure the goodness of a model fit to data.
A typical metric is $\chi^2=\sum_{j=1}^{N} (f_p({x_{j}}) - y_{j})^2$ which measure the goodness of model fitting to data.
Minimizing a series of metrics achieves a reasonable solution after many iterations 
with the numerical gradient descent method given a learning rate.
This technique is called symbolic regression~\cite{paper:symbreg}.
Symbolic regressions not only find models but also solve them, while traditional regressions only solve a model for a given data.
The return values of metrics are used to guide the evolution direction.
For examples, models like $p_1/Y$ or $p_1/I$ will be kicked out at the beginning if data favors $p_1Y$.
Generally speaking, traditional approaches will stop here, 
but this approach will further guide the evolution of current model.

\item Generalization:
the generalization proceeds along two directions, the model and the data.
For the former case, we need an invariant model.
For example, the gravitation law works for all planets in the solar system.
For the later case, we need a model work well for all data at different domain. 
For example, relativity theory works for motion objects of both low velocity and high velocity.
Generalization possibly bring new knowledge as discussed in next step. 

\item Inference: 
In case of holding the form of current model invariant, refinement of current model with updated data
can figure out connections among free parameters.
In case of expanding current model,  new evolution of current model can tell hidden concepts. 
For example, suppose a model $F=p_1/r^2$ be learned from the trajectory data of Earth, 
a new model $F=p_2/r^2$ can be obtained from the trajectory data of Mars.
Generalization results in a better model $F=p_{12}m_i/r^2$ 
and infers a planet-related concept $m_i$ (i=Earth,Mars) which actually represents a dimensionless mass normalized by the mass of Earth~\cite{PhysRevLett.124.010508}. 

\item Output: 
current model will be stored in model database for future prediction, unification and expansion. 
\end{enumerate}

In a word, this framework combines the basic ideas of physics with guidelines of machine learning to build a reason system, by which hidden equations or new physics concepts can be inferred.

Time went back in 1960s, newly discovered hadrons led to Enrico Fermi said "Young man, if I could remember the names of these particles, I would have been a botanist."  A classification schemes for hadrons became as natural as what Mendeleev did for chemical elements.
Starting from one of fundamental concepts symmetry, Gell-Mann and Okubo have actually studied the relation of hadron mass, hypercharge and isospin of baryon decuplet and octet. They obtained a formula
\begin{equation}\label{eq:Gellmann-Okubo}
M = a + bY + c[I(I+1) - \frac{1}{4}Y^2]
\end{equation}
with prior knowledge of particle physics.
Parameters ($a$, $b$, $c$) are  extracted from measured hadron masses for a given irreducible representation. From Eq.~\eqref{eq:Gellmann-Okubo}, one can see that hadrons' mass ($M$) can be determined by their quantum numbers, i.e. 
hypercharge ($Y$), isospin ($I$), which stems from the underlying dynamics. Thus, we use the machine-learning-based approach to discover the Gell-Mann–Okubo formula from the SU(3) baryon decuplet and octet.
\begin{table}[h]
    \centering
        \caption{The dataset of decuplet ($J^P=\frac{3}{2}^+$) and octet ($J^P=\frac{1}{2}^+$) baryons. 
        The masses in the first column are taken from Ref.~\cite{Workman:2022ynf}. The central values are the isospin averaged ones. Their uncertainties are set to cover all the masses within the same isospin multiplet instead of the experimental values.}
    \begin{tabular}{c|c |c |c |c}
    \hline\hline
          Y & I & $J^{P}=\frac{1}{2}^{+}$ & $J^{P}=\frac{3}{2}^{+}$ &Mass (MeV/c$^2$) \\
         \hline
          1 & $\frac{1}{2}$ & $n,p$ & & 939$\pm$1  \\
          0 & 0 & $\Lambda$ & &1116$\pm$1 \\
          0 & 1 & $\Sigma^{-}, \Sigma^{0}, \Sigma^{+}$ & &1193$\pm$4 \\
          -1 & $\frac{1}{2}$ & $\Xi^{-}, \Xi^{0}$ & &1318$\pm$3 \\
          1 & $\frac{3}{2}$ & & $\Delta^{-}, \Delta^{0}, \Delta^{+}, \Delta^{++}$&1232$\pm$2 \\
          0 & 1 & & $\Sigma^{*-}, \Sigma^{*0}, \Sigma^{*+}$&1385$\pm$3 \\
          -1 & $\frac{1}{2}$ & & $\Xi^{*-}, \Xi^{*0}$&1533$\pm$2 \\
          -2 & 0 & & $\Omega^{-}$&1672$\pm$1 \\
         \hline
    \end{tabular}
    \label{tab:massdata}
\end{table}

\vspace{0.2cm}
{\bf \color{gray}Results and Discussions:}
With above framework, we show an example in discovering the Gell-Mann–Okubo formula Eq.~\eqref{eq:Gellmann-Okubo} from baryon spectrum.
As the concepts of isospin and hypercharge had been well established before the Gell-Mann-Okubo formula, 
we directly start from observable ${\bm{x}}=(Y,I)$ and the target observable ${\bm{y}}=(M)$ listed in Tab.~\ref{tab:massdata}. 
A predefined set of above observable $\{M,~Y,~I\}$ and operators $\{+,~-,~*,~/\}$
as listed in Tab.~\ref{tab:opobs} is used to build models $f_C(Y,I)$ 
where C are free parameters. 
An evolution means that a model $f_C(Y,I)$ changes into a different form $f'_{C'}(Y,I)$.
Parameters belonging to the model $f_C(Y,I)$ are solved with the gradient descent method after $N$ times iteration,
\begin{eqnarray}\label{eq:chisq}
\chi^2 & = & \sum_{k}^{N} \frac{(f_C(Y,I) - M)_k^2}{\epsilon^2_k} \\
\label{eq:gradient descent}
C_j^{i+1} & = &C_j^{i}-\alpha*\frac{\partial{\chi^2}}{\partial{C_j^i}}
\end{eqnarray}
where $\alpha$ is the learning rate, $\epsilon_k$ is the $k$th particle's mass error, $C_j^i$ is the $j$th parameter for the $i$th iteration and $C_j^{i+1}$ is the updated value for the $(i+1)$th iteration. $\frac{\partial{\chi^2}}{\partial{C_j}}$ is the gradient of $\chi^2$ with respect to $C_j$.

\begin{figure*}[h]
    \centering
    \includegraphics[width=0.8\textwidth]{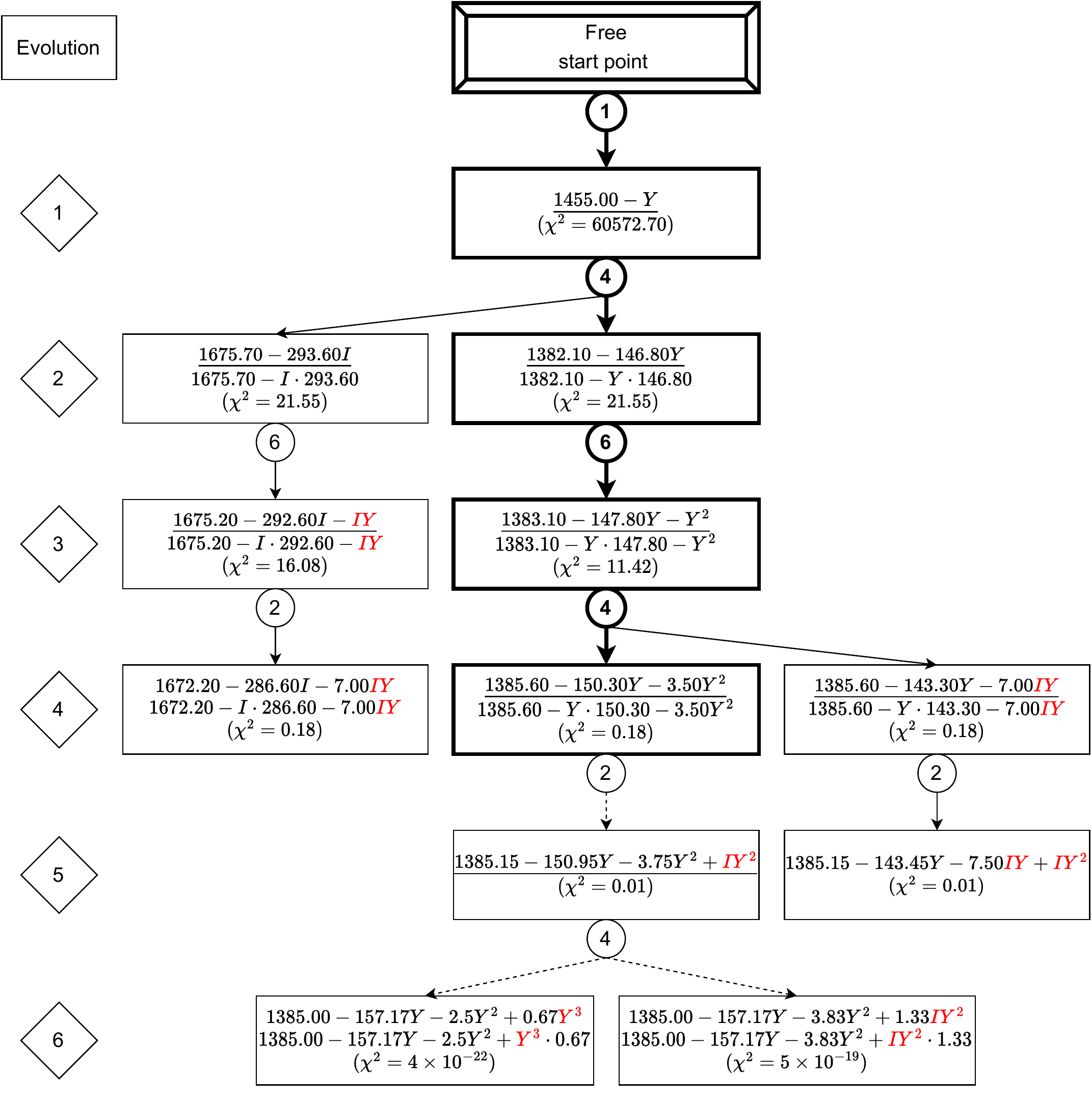}
    \caption{Evolution results with baryon decuplet.
    The first column represent the times of evolution. 
    The equations in the rectangle represent the evolutionary results of the generation, where the underlined equations are those that need to evolve, and the numbers in the circles below the rectangle represent the number of evolutionary results of the next generation. Besides, the bolded and dashed lines represent the best and ineffective evolutionary processes, respectively.}
    \label{fig:decuplet}
\end{figure*}

We firstly test data of the SU(3) flavor symmetric decuplet.
At the beginning, enumeration of the simplest models consisting of one operator and two observable 
yields $\{C_1Y, C_1I, C_1+Y,C_1+I, C_1/Y, C_1/I,C_1-Y,C_1-I,I+Y,I-Y,IY,I/Y...\}$,
among which the model $\{C_1-Y\}$ is favourite.
Next, it evolves to either one branch $A:\{C_1-C_2I\}$ (the left column in Fig.~\ref{fig:decuplet}) or the other branch $B:\{C_1-C_2Y\}$ (the middle column in Fig.~\ref{fig:decuplet}).
Evolution of branch A reaches a solution $\{C_1-IY+C_2I\}$ with a reasonable $\chi^2$ value as listed in Tab.~\ref{tab:decuplet}.  
However, this solution is rejected since the $IY$ item is prohibited as isospin and hypercharge are different degrees of freedom.
Evolution of branch B reaches a solution $\{C_1-C_2Y-C_3Y^2\}$ and another solution $C:\{C_1-C_2Y-C_3IY\}$ (the right column in Fig.~\ref{fig:decuplet}).
Note that, more evolution of solution B raises items like $IY^2$ or $Y^3$ and corresponding $\chi^2$ value becomes extremely small, which indicates a termination sign of over-fitting. As the result, we obtain an expression for baryon decuplet,
\begin{equation}\label{eq:ten}
M_{\mathrm{decuplet}}=C_1-C_2Y-C_3Y^2.
\end{equation}

\begin{table}[h]
    \centering
        \caption{Evolution results for baryon decuplet.
        The first and second columns represent the times of evolution and the corresponding $\chi^2$, respectively. 
        The right column is the evolution result for each evolution time with the explicit values of all $C_i$s.
        A learning rate $\alpha=0.01$ is used. Branch A, B, C correspond to the left, the middle and the right column of Fig.~\ref{fig:decuplet}, respectively. The bold formula is accepted,  while formulas with underlined items  are rejected by physics constraints. }
    \begin{tabular}{c|c |c }
    \hline\hline
    \multicolumn{3}{c}{Branch A}\\
    \hline
         Evolution & $\chi^2$ & Model  \\
         \hline
         1 &  60572.70 &   $1455.00-Y$   \\
         2 &  21.55    &  $1675.70-293.60I$ \\
         3 &  16.08    &$1675.20-292.60I-\uline{IY}$ \\
         4 &  0.18     & $1672.20-286.60I-7.00\uline{IY}$ \\
         \hline
    \hline
    \multicolumn{3}{c}{Branch B}\\
    \hline
         Evolution & $\chi^2$ & Model  \\
         \hline
         1 &  60572.70 &    $1455.00-Y$   \\
         2 &  21.55    & $1382.10-146.80Y$\\
         3 &  11.42 &  $1383.10-147.80Y-Y^2$ \\
         4 &  0.18   & $\bm{1385.60-150.30Y-3.50Y^2}$ \\
         5 &  0.01 & $1385.15-150.95Y-3.75Y^2+\uline{IY^2}$ \\
         6a &  $5*10^{-19}$ & $1385.00-151.17Y-3.83Y^2+1.33\uline{IY^2}$\\
         6b &  $4*10^{-22}$ & $1385.00-151.17Y-2.50Y^2+0.67\uline{Y^3}$\\
     \hline
    \hline
    \multicolumn{3}{c}{Branch C}\\
    \hline
         Evolution & $\chi^2$ & Model  \\
         \hline
         1 &  60572.70 &    $1455.00-Y$   \\
         2 &  21.55    & $1382.10-146.80Y$\\
         3 &  11.42 &  $1383.10-147.80Y-Y^2$ \\
         4 &  0.18   & $1385.60-143.30Y-7.00\uline{IY}$ \\
         5 &  0.01 & $1385.15-143.45Y-7.50\uline{IY}+\uline{IY^2}$ \\
     \hline
    \end{tabular}  \label{tab:decuplet}
\end{table}
\begin{figure*}[h]
    \centering
    \includegraphics[width=0.8\textwidth]{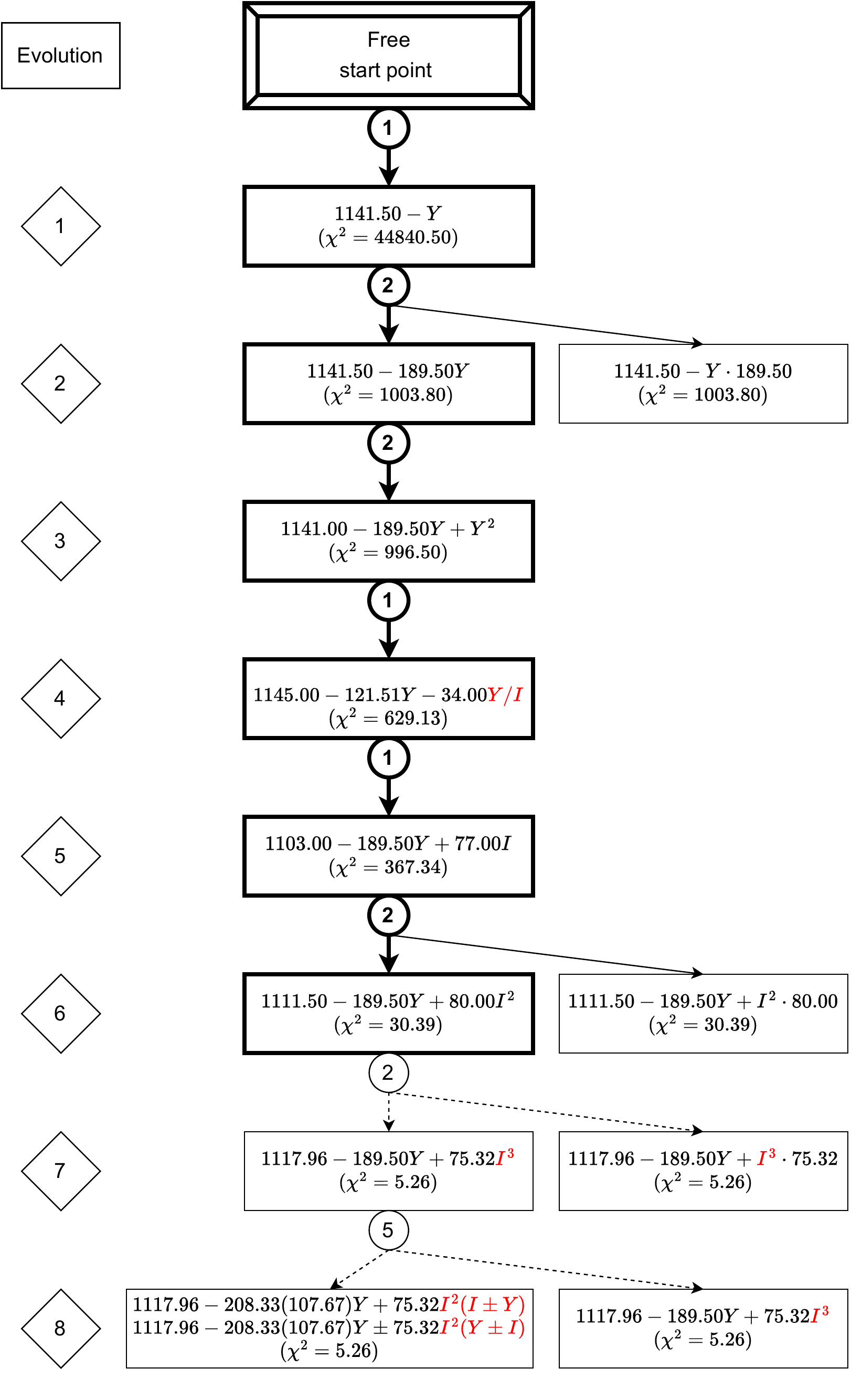}
    \caption{Evolution results with baryon octet. The labels are analogous to that of Fig.~\ref{fig:decuplet}}
    \label{fig:octet}
\end{figure*}
We perform a new evolution using baryon octet,
and list results in Tab.~\ref{tab:eight} with repeating aforementioned procedure.
As the result, we obtain an expression for baryon octet,
\begin{equation}\label{eq:eight}
M_{\mathrm{octet}}=C_a-C_bY+C_cI^2.
\end{equation}
\begin{table}[h]
    \centering
        \caption{Evolution results for baryon octet. The caption is analogous to that of Tab.~\ref{tab:decuplet}.}
    \begin{tabular}{c|c|c}
    \hline\hline
         Evolution & $\chi^2$ & Model \\
         \hline
         1 &  44840.50  &   $1141.50-Y$    \\
         2 &  1003.80  & $1141.50-189.50Y$ \\
         3 &  996.50  & $1141.00-189.50Y+Y^2$ \\
         4 &  629.13  & $1145.00-121.51Y-34.00\uline{Y/I}$\\
         5 &  367.34  &$1103.00-189.50Y+77.00I$  \\
         6 &  30.39 & $\bm{1111.50-189.50Y+80.00I^2}$\\
         7 &  5.26 & $1117.96-189.50Y+75.32\uline{I^3}$ \\ 
         8a &  5.26 &  $1117.96-189.50Y+75.32\uline{I^3}$ \\ 
         8b &  5.26 & $1117.96-208.33Y+75.32\uline{I^2(I+Y)}$\\
         8c &  5.26 & $1117.96-170.67Y+75.32\uline{I^2(I-Y)}$\\
         \hline
    \end{tabular}
    \label{tab:eight}
\end{table}

\begin{widetext}
\begin{center}
\begin{table}[h]
    \centering
        \caption{Evolution results after applying the generalization to two models for baryon decuplet and octet. The first and second columns are the evolution time and the corresponding $\chi^2$. The further evolution starts from the expression of baryon decuplet, i.e.Eq.~\ref{eq:ten}. The third and forth columns are the explicit results for baryon decuplet and octet, respectively. The bold expressions are the final results. The terms underlined are rejected by physical constraints. }
    \begin{tabular}{c|c|c|c}
    \hline\hline
         Evolution & $\chi^2_8+\chi^2_{10}$&Decuplet &Octet \\
         \hline
         1 &  0.2055520 &$803.16-441.52Y-3.50Y^2+582.44I$ & $1115.84-189.50Y-25.95Y^2+77.24I$\\
         2 &  0.1800060 &$511.97-587.12Y-3.50Y^2+873.63I$  & $1116.00-189.50Y-26.00Y^2+77.00I$\\
         3 & 0.1799985 &$\bm{689.54-672.34Y-90.51Y^2+348.03I(I+1)}$& $\bm{1116.00-189.50Y-16.38Y^2+38.50I(I+1)}$\\
         4 &  0.1208117 &$1381.94-153.19Y-4.12Y^2+2.48\uline{(I(Y^2}+1))$ & $1116.00-189.50Y-64.50Y^2+77.00\uline{(I(Y^2}+1))$\\
         5 &  $1*10^{-10}$ &$1384.33-151.50Y-3.83Y^2+0.67I+1.33\uline{IY^2}$ & $1116.00-189.50Y-103.00Y^2+77.00I+154.00\uline{IY^2}$\\
         \hline
    \end{tabular}
    \label{tab:baryon}
\end{table}
\end{center}
\end{widetext}

A further evolution required by generalization
is possible to establish a universal equation of incorporating both Eq.~\eqref{eq:ten} and Eq.~\eqref{eq:eight}
using all data of baryon decuplet and octet.
This evolution starts from Eq.~\eqref{eq:ten},
and is simultaneously guided by two metrics $\chi^2_8$ and $\chi^2_{10}$, i.e. the $\chi^2$ defined in Eq.~\eqref{eq:chisq} for baryon octet and decuplet. 
The first metric $\chi^2_8$ makes sure a model work well for baryon octet,
and the second metric $\chi^2_{10}$ makes sure a copy of current model work well for baryon decuplet.
Note that the form of current model is hold while its parameters are free.  
Evolution results are shown in Tab.~\ref{tab:baryon}.
A generalized model for both baryon decuplet and octet can be taken as either
\begin{equation}\label{eq:mass1}
M=C_1-C_2Y-C_3Y^2+C_4I
\end{equation}
or 
\begin{equation}\label{eq:mass2}
M=C_1-C_2Y-C_3Y^2+C_4(I(I+1)).
\end{equation}
In both cases, the $\chi^2$ values are reasonable.
Furthermore, without the unkonwn SU(3) flavory symmetry breaking mechanism, for instance the electromagnetic interaction, particles within the same isospin multiplet should have the same mass. The value of the mass should be the eigen value of the operator $\hat{I}^2$, i.e. $I(I+1)$. Thus, the formula $C_1-C_2Y-C_3Y^2+C_4(I(I+1))$ 
can be accepted by physics. The number of the parameters 
is consistent with that in the octet model~\cite{deSwart:1963pdg}, where the four free parameters are the reduced matrix elements of the irreducible representations of one singlet, two octets and one 27-plet from the group theory point of view. 
The Gell-Mann–Okubo mass formula, i.e. Eq.\eqref{eq:Gellmann-Okubo},
is obtained based on the assumption that the reduced matrix element of the 27-plet
is zero. Thus the number of the free parameters is reduced to three. 
In our framework, this relation can also be seen by the relations between $Y$ and $I$ hidden in baryon decuplet (Eq,\eqref{eq:tenYIa}) and octet (Eq,\eqref{eq:eightYIb}) after evolution as listed in Tab.~\ref{tab:eightYI}.
\begin{eqnarray}
Y & = & 2I-2, \ \text{for baryon decuplet} \label{eq:tenYIa} \\
Y^2& = & 4I-4I^2, \ \text{for baryon octet} \label{eq:eightYIb}
\end{eqnarray}
Substituting Eq.~\eqref{eq:eightYIb} to Eq.~\eqref{eq:mass2},
one obtain
\begin{equation}
M=C_1-C_2Y+(C_4-4C_3) I+(C_4+4C_3)I^2.
\end{equation}
When $C_4=4C_3$, the above formula comes back to 
the mass formula Eq.~\eqref{eq:eight} evolution from baryon octet. Along the same line, one can substitute Eq.~\eqref{eq:tenYIa}
to Eq.~\eqref{eq:mass2} and obtain
\begin{equation}
M=C_1+2C_4-(C_2-\frac{3C_4}{2})Y+(\frac{C_4}{4}-C_3) Y^2. 
\end{equation}
Analogously, when $C_4=4C_3$, the above formula comes back to the expression
$C_1+2C_4-(C_2-\frac{3C_4}{2})Y$. Although the expression
does not come back to Eq.~\eqref{eq:ten} as we expected, 
the vanishing $Y^2$ term can be obviously seen by the final result in Tab.~\ref{tab:decuplet}, i.e. the coefficient of $Y^2$ is about two orders smaller than the others. 
As discussion above, to obtain an universial mass formula for both baryon decuplet and octet, $C_4=4C_3$ is required, which leads Eq.~\eqref{eq:mass2} to the Gell-Mann-Okubo mass formula Eq.~\eqref{eq:Gellmann-Okubo}.

\begin{table}[h]
    \centering
        \caption{Evolution results between $Y$ and $I$ for baryon decuplet and octet. Note that $Y$ and $I$ have no errors.
         A learning rate  $\alpha=0.01$ is used. The first and second columns are for the evolution time and the corresponding $\chi^2$. The last column is the evolution result. The bold formulae are final results. }
    \begin{tabular}{c|c |c }
    \hline\hline
    \multicolumn{3}{c}{baryon decuplet}\\
    \hline
         Evolution & $\chi^2$ & Model \\
         \hline
         1 &  1.25  &      $I-1.25$  \\
         2 &  $5*10^{-16}$    & $\bm{2.00I-2.00}$ \\
         \hline
\hline
    \multicolumn{3}{c}{baryon octet: Branch A}\\
    \hline
         Evolution & $\chi^2$ & Model \\
         \hline
         1 &  1.00  &  $0.50+0.50$ \\
         2 &  0.56     &  $0.375+I-I^2$ \\
         \hline
\hline
    \multicolumn{3}{c}{baryon octet: Branch B}\\
    \hline
         Evolution & $\chi^2$ & Model \\
         \hline
         1 &  1.00  &  $0.71*0.71$ \\         
         2 &  $1*10^{-15}$ & $\bm{4.00(I-I^2)}$ \\
         \hline
    \end{tabular}
    \label{tab:eightYI}    
\end{table}

\vspace{0.2cm}
{\bf \color{gray} Summary and Outlook:}
Machine learning is a novel and powerful technology
for commerce and industry. It has been widely and successfully used in various 
science topics~\cite{paper:symm}~\cite{paper:gravitation}.
However, its application in hadron physics, especially theoretical hadron physics,
is still on its early stage. Analogous to the role of periodic table of elements in chemistry, hadron spectrocopy can also imply
the underlying dynamics. As the result,
we start from hadron spectroscopy, more specifically 
baryon decuplet and the baryon octet. 
A physics-inspired machine learning approach
is constructed in this work to
 discover the Gell-Mann–Okubo formula. 
This approach can also be used to study the mass formula for exotic hadrons discovered and intensely discussed recently years. 

\vspace{0.2cm}
{\bf \color{gray}Acknowledgement}
This work is partly supported by Guangdong Major Project of Basic and Applied Basic Research No.~2020B0301030008,
the National Natural Science Foundation of China with Grant No.~12035007, Guangdong Provincial funding with Grant No.~2019QN01X172.
Q.W. is also supported by the NSFC and the Deutsche Forschungsgemeinschaft (DFG, German
Research Foundation) through the funds provided to the Sino-German Collaborative
Research Center TRR110 ``Symmetries and the Emergence of Structure in QCD"
(NSFC Grant No. 12070131001, DFG Project-ID 196253076-TRR 110).

\nocite{*}
\bibliography{ref.bib}

\end{document}